\documentclass[12pt]{iopart}
\usepackage{graphicx}
\usepackage{bm,amssymb,citesort}

\renewcommand{\vec}[1]{\mathbf{#1}}
\newcommand{\EE}{{\cal E}}

\begin{document}
\title{The Glassy Wormlike Chain}
\author{Klaus Kroy$^{1,2}$ and Jens Glaser$^1$}
\address{$^1$ Institut f\"ur Theoretische Physik, Universit\"at Leipzig,
Postfach 100920, 04009 Leipzig, Germany}
\address{$^2$ Hahn-Meitner-Institut, Glienicker Stra\ss e 100, 14109 Berlin,
Germany}
\eads{klaus.kroy@itp.uni-leipzig.de}

\begin{abstract}
We introduce a new model for the dynamics of a wormlike chain in an
environment that gives rise to a rough free energy landscape, which we
baptise the glassy wormlike chain. It is obtained from the common
wormlike chain by an exponential stretching of the relaxation spectrum
of its long-wavelength eigenmodes, controlled by a single parameter
$\EE$.  Predictions for pertinent observables such as the dynamic
structure factor and the microrheological susceptibility exhibit the
characteristics of soft glassy rheology and compare favourably with
experimental data for reconstituted cytoskeletal networks and live
cells. We speculate about the possible microscopic origin of the
stretching, implications for the nonlinear rheology, and the potential
physiological significance of our results.
\end{abstract}
\submitto{\NJP}



\section{Introduction}
Studies of \emph{in vitro} polymerised networks and solutions of the
biopolymers that constitute the cytoskeleton have provided many
important insights into the molecular origin of the fascinating
mechanical properties of cells and tissues \cite{Bausch2006}.  Cell
rheological data \cite{wottawah2005,Fernandez2006} are therefore often
interpreted in terms of viscoelastic models developed for \emph{in
vitro} reconstituted cytoskeletal networks
\cite{Morse2001,Gardel2004Science,Storm2005,Liu2006,tharmann2007}, and
the inferred parameters have been suggested as pertinent indices for
clinical diagnosis \cite{Guck2005}. Also current rheological models
for the active processes in the cytoskeleton are rooted in this
viscoelastic paradigm
\cite{hatwalne-etal:2004,kruse-etal:2004,mizuno-tardin-schmidt-mackintosh:2007}. In
contrast, microrheological measurements \cite{Fabry2001} have revealed
that cells obey the highly universal and comparatively featureless
pattern of soft glassy rheology \cite{Sollich1997}, ubiquitous in soft
condensed matter \cite{cipelletti-ramos:2005}. Integrating these
competing paradigms into a unified framework has become a major
challenge in cell biophysics
\cite{Deng2006,Hoffman2006,Rosenblatt-etal:2006}. Recent experiments
with highly purified reconstituted actin solutions have revealed
strong signatures of a glass transition that might hold the key to a
resolution of the dilemma \cite{semmrich-etal:tbp}. The glass
transition manifests itself in a strong stretching of the relaxation
spectrum, so that experiments probing the mechanical properties at a
fixed time scale detect a sharp transition from fluid-like to
solid-like behaviour as a function of various control parameters. In
the following, we introduce a new model motivated by this observation,
which is able to explain the apparently conflicting phenomenology on a
common basis. We call it the glassy wormlike chain (GWLC), because its
essence is an exponential stretching of the relaxation spectrum of the
wormlike chain (WLC) model, which is the minimal model of a
semiflexible polymer. The stretching is quantified by a single
parameter $\EE$, the stretching parameter, which can be thought of as
a characteristic scale for the free energy barriers retarding the
relaxation of the test chain's long wavelength eigenmodes.  The
characteristics of soft glassy rheology naturally ensue, including the
widely reported small apparent power-law exponents in the
frequency-dependence of the microrheological moduli
\cite{Fabry2001,Sollich1997,Deng2006,Hoffman2006}, or the so-called
``noise temperatures''.  Despite of the striking simplicity of the
model ($\EE$ is the only free parameter), which is in stark contrast
to prevailing viscoelastic models for cytosekeletal networks \emph{in
vitro} and \emph{in vivo} \cite{Morse2001,ananthakrishnan-etal:2005},
its predictions compare favourably with experimental data for live
cells and reconstituted cytoskeletal networks. In particular, the
logarithmic tails of the dynamic structure factor obtained by
high-precision quasi-elastic light scattering from F-actin solutions
at low temperatures and high concentrations \cite{semmrich-etal:tbp}
provide very direct evidence for the postulated exponential
stretching. While a microscopic derivation of $\EE$ remains so far
elusive, we can demonstrate the practical usefulness of the model for
many applications by computing pertinent measurable quantities.

\section{The wormlike chain (WLC)}\label{sec:WLC}
In the WLC model a semiflexible polymer is represented as a continuous
space curve $\vec r(s)=(\vec r_\perp(s), s-r_\|(s))$ with arc length
$s=0\dots L$.  We consider the weakly-bending rod limit where
deflections $\vec r_\perp'(s)$ from the straight ground state are
considered small, $\vec r_\perp'(s) \ll 1$, the prime denoting an
arc length derivative.  From the arc length constraint, ${\vec
r'}^2(s)=1$, it follows that longitudinal fluctuations are of higher
order, $r_\|'={\cal O}({\mathbf r'}_\perp^2)$. The dynamics of a
weakly bending WLC subject to an (optional) constant backbone tension
$f$ is to leading order described by the linear Langevin equation for
its transverse excursions $\vec r_\perp(s,t)$
\begin{equation}\label{eq:langevin}
\zeta_\perp \dot{\vec r}_\perp = -\kappa \vec r_\perp''''+ f \vec
    r_\perp''+ \boldsymbol \xi_\perp \;.
\end{equation}
Here $\kappa$, $\zeta_\perp$ and $\boldsymbol\xi_\perp(s,t)$ denote
the bending rigidity, the solvent friction per length, and Gaussian
thermal noise, respectively. For more details about WLC dynamics the
reader is referred to
Ref.~\cite{hallatschek-frey-kroy:2007a}. Equation~(\ref{eq:langevin})
is solved by introducing eigenmodes,
\begin{equation}
\vec r_\perp(s,t) = \sum_{n=1}^\infty {\mathbf a}_n(t)
W_n(s)\,.
\end{equation}
For simplicity we assume hinged boundary conditions, in which case the
eigenfunctions are simple sine functions
\begin{equation}
W_n(s) = \sqrt{\frac{2}{L}} \sin(k_n s)
\end{equation}
and the eigenvalues $k_n= n \pi/L$ can be parametrised by natural
numbers $n$.  The eigenmodes relax independently and exponentially
\begin{equation}
\langle {\mathbf a}_n(t) {\mathbf a}_m(0) \rangle = \delta_{nm}\langle {\mathbf a}_n^2 
\rangle \exp(-t/\tau_n)\;.
\end{equation}
The equilibrium mode amplitudes
\begin{equation}
\langle {\mathbf a}_n^2 \rangle = \frac{2k_BT}{\kappa k_n^4+f k_n^2}
\end{equation}
follow from equipartition and the mode relaxation time is given by
\begin{equation}
\tau_n = \tau_L/(n^4 + n^2f/f_L)
\end{equation}
with the relaxation time $\tau_L= \zeta_\perp L^4/\kappa\pi^4$ and the
Euler force $f_L= \kappa \pi^2/L^2$ of the longest mode
setting the characteristic time and force scale, respectively.

These results can be used to calculate various time-dependent
correlation functions as a superposition of eigenmode contributions.
For example, the transverse dynamic mean-square displacement (MSD)
reads
\begin{equation}\label{eq:MSD}
\delta r_{\perp L}^2(t)\equiv \langle(\vec r_{\perp}(s,t)-\vec r_{\perp
}(s,0))^2 \rangle=\frac{4L^3}{\ell_p \pi^4} \sum_n \frac{1-\exp(-
t/\tau_n)}{n^4 + n^2f/f_L}\;.
\end{equation}
The MSD is directly or indirectly measured by a couple of experimental
techniques, especially by particle tracking, dynamic light scattering,
and diverse passive and (linear) active microrheology methods.

\section{The glassy wormlike chain (GWLC)}
The GWLC model is obtained from the WLC by an exponential stretching
of the relaxation spectrum in the spirit of so-called hierarchically
constrained dynamical models \cite{brey-prados:2001}. The strategy is
also reminiscent of the generic trap models \cite{Monthus1996}
underlying soft glassy rheology \cite{Sollich1997}, but concerns the
equilibrium dynamics of the test chain, here. The GWLC is a WLC with
the relaxation times for all its eigenmodes of mode number $n<l\equiv
L/\Lambda$ --- or, more intuitively, of (half) wavelength
$\lambda_n\equiv L/n > \Lambda$ --- modified according to
\begin{equation}\label{eq:GWLC}
\tau_n \to \tilde \tau_{n} =
\Bigg\{
\begin{array}{cl}
\tau_{n} & (n>l) \\
\tau_n \exp(N_n\EE ) & (n<l) \;
\end{array}.
\end{equation}
Here
\begin{equation}\label{eq:N}
N_n \equiv l/n-1= \lambda_n/\Lambda -1 
\end{equation}
can be thought of as the number of interactions per length $\lambda_n$
with the environment and $\Lambda\ll \ell_p,\;L$ as a characteristic
interaction length. We moreover introduce the suggestive notation
$\tau_\Lambda\equiv \tau_l=\zeta_\perp\Lambda^4/\kappa\pi^4$ and
$\omega_\Lambda =2\pi/\tau_\Lambda$ for the corresponding crossover
time and frequency, respectively. We imagine the retardation of the
relaxation of the long-wavelength modes of the test polymer to be
caused by a complex environment such as a solution of other
polymers. For example, in a semidilute solution of semiflexible
polymers, $\Lambda$ would correspond to the familiar entanglement
length \cite{semenov:86}, $\omega_\Lambda$ to the entanglement
frequency, and $N_n$ to the number of entanglements an undulation of
arc length $\lambda_n$ has to overcome in order to relax.  With regard
to the application to cytoskeletal networks we have in particular
polymers with incompletely screened sticky interactions in mind (see
below). The parameter $\EE$, which we call the \emph{stretching
parameter}, controls the slowing down caused by the interactions. It
is the key parameter of the model --- and in fact the only parameter
apart from the rather obvious interaction scale $\Lambda$. Physically,
it may be interpreted as a characteristic height of the free energy
barriers in units of thermal energy $k_BT$ in a rough free energy
landscape.  Below, we speculate about the microscopic origin of such a
free energy landscape in cytoskeletal networks and try to give
tentative estimates for various contributions to $\EE$.

Our definition of the GWLC does only affect the relaxation times but
not the amplitudes $\vec a_n$ of the test chain's eigenmodes. The
question, how the amplitudes of the long wavelength modes are affected
by the interactions with the disordered environment is in principle an
interesting open question that is currently under
investigation. However, we expect that the slowing-down captures the
most important mechanism underlying the observed glass transition and
the corresponding soft glassy rheology, and that the conformational
aspects are in this respect of minor relevance.

\begin{figure}
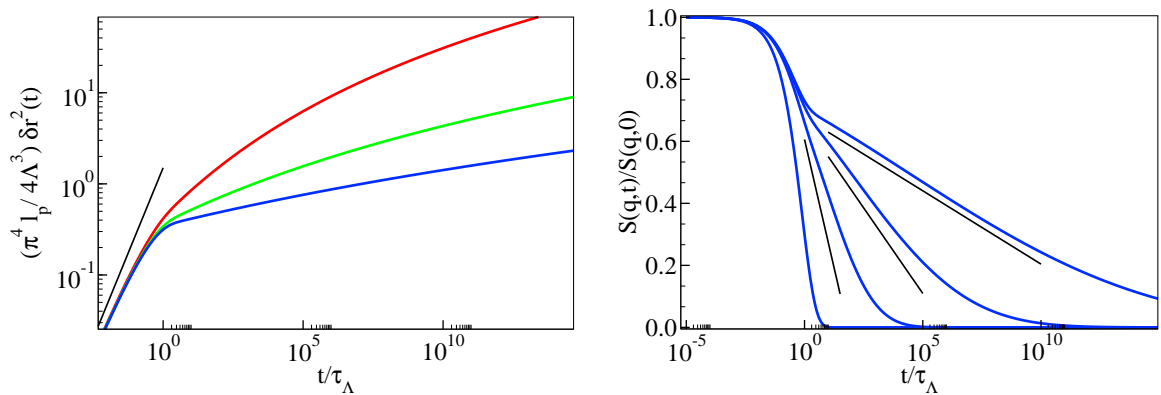

\begin{center}
\includegraphics[width=0.48\columnwidth]{msd.eps} \quad
\includegraphics[width=0.46\columnwidth]{sf.eps}
\end{center}
\caption{Time dependence of the transverse dynamic mean-square
  displacement $\delta r_{\perp L}^2(t)$ (MSD) and the dynamic
  structure factor $S(q,t)$. \emph{Left:} The normalised MSD of a GWLC
  as a function of time $t$ for stretching parameters $\EE=5$, 15, 35
  (from top to bottom) for $f=0$. The straight line indicates the limiting
  power-law growth $\delta r_{\perp L}^2(t) \propto t^{3/4}$ of the
  MSD of an infinite weakly bending WLC. \emph{Right:} The dynamic
  structure factor of a GWLC, evaluated numerically for $q^2=\ell_p
  \pi^4/\Lambda^3$ and stretching parameters ${\cal E}=0, 5, 15, 35$
  (from left to right). The straight lines indicate the logarithmic
  intermediate asymptotics calculated in
  equation~(\ref{eq:Sqt2}).}\label{fig:Sqt}
\end{figure}

As a first example of a pertinent observable for the GWLC, we plot in
figure~\ref{fig:Sqt} (left) the MSD for various values of the
stretching parameter at vanishing prestress.

\section{Prestress}
Before we come to the evaluation of further observables, we first want to
consider the effect of tension on a GWLC. An (optional) constant
tension $f$ was already included in our brief account of the ordinary
WLC in section~\ref{sec:WLC}. However, we should certainly also expect
an effect of any kind of external or internal stress onto the escape
of the polymer over the free energy barriers represented by
$\EE$. Intuitively, the force is expected to ``help the polymer over
the free energy barriers'', but we cannot, of course, exclude the
opposite effect, namely that the traps become under certain
circumstances deeper upon applying a force. In any case, the natural
way to introduce a force into this picture is via a tilting of the
free energy landscape in the spirit of a Kramers escape rate model,
i.e.
\begin{equation}\label{eq:kramers}
 \EE \to \EE \pm f/f_T\;, \qquad  f_T\equiv k_BT/\Delta \;.
\end{equation}
The minus sign corresponds to the force lowering the barrier. The
length $\Delta$ should be interpreted as a characteristic width of the
free energy wells and barriers. Accordingly, $f_T$ represents the
scale of thermally induced force fluctuations, which are even present
in absence of an applied stress.

The introduction of an external force in equation~(\ref{eq:kramers})
may be seen as a simple heuristic method to address the rheology of
prestressed networks or even of the nonlinear rheology of cytoskeletal
networks. This is of considerable interest for potential applications
of our model, since prestress is thought to be the crucial element
needed for mimicking typical cell rheological behaviour using much
simpler reconstituted networks
\cite{wang-etal:2002,ingber:2003,lau-etal:2003,Gardel2006PNAS,Rosenblatt-etal:2006}.
There is, however, a subtle point involved in the interpretation of
the force $f$ as arising from a prestress. For simplicity, we have
tacitly assumed above that the force that pulls the test chain over
the free energy barriers is identical to its backbone tension. This
should not \emph{a priori} be a problematic assumption for qualitative
purposes. But a test polymer embedded into a cytoskeletal network or
sticky biopolymer solution that was initially unstressed will
generally have a conformation different from the stretched
conformation of our free test chain equilibrated under the backbone
tension $f$. As noted above, we do not attempt to address this
problem, at the present stage.  Altogether, after disregarding such
potential modifications of the equilibrium conformation of the test
chain by its surroundings and identifying the barrier lowering force
with the backbone tension, $f$ remains to be related to the
macroscopic shear stress. Consistent with our discussion of the shear
modulus in section~\ref{sec:modulus}, below, we follow
Ref.~\cite{gittes-mackintosh:98_pub} in writing $\sigma=f/5\xi^2$,
where $\xi\equiv (3/c_pL)^{1/2}$ is the mesh size of a semidilute
solution of semiflexible polymers. Its relation to the polymer
concentration $c_p$ and the numerical prefactors are of geometric
origin. As a reminder of the tentative nature of the identification of
$\sigma$ as an actual prestress we call $\sigma$ the ``nominal
prestress''.

\section{Dynamic structure factor}

For sufficiently long times $t\gg \zeta_\perp/\kappa q^4$, the dynamic
structure factor $S(q,t)$ at scattering vector $q\gg\Lambda^{-1}$
follows from the transverse mean square displacement according to
\cite{kroy-frey:2000},
\begin{equation}\label{eq:Sqt}
S(q,t)/S(q,0) \sim \exp\left[-q^2 \delta r_{\perp
L}^2(t)/4\right]\;.
\end{equation}
The time dependence of the structure factor is plotted in
figure~\ref{fig:Sqt}. A pronounced logarithmic intermediate
asymptotics is seen to develop for large $\EE$.  It can be calculated
approximately as follows. First, we identify a logarithmic
intermediate asymptotics in the MSD. At intermediate times the MSD
takes the form
\begin{equation}\label{eq:logtail}
\delta r_{\perp L}^2(t) \sim \delta r_{\perp\Lambda}^2(\infty)+
\frac{4\Lambda^3}{\EE \ell_p \pi^4} \left[\gamma_E +
\log(t/\tau_\Lambda)\right] \qquad (1 \ll t/\tau_\Lambda \ll \EE)\;,
\end{equation}
for an unstressed chain ($f=0$). Here $\delta
r_{\perp\Lambda}^2(\infty)$ contains the saturated contributions from
the free WLC modes up to wavelength $\Lambda$. The GWLC part of the
mode spectrum is asymptotically approximated by the logarithmic term.
Expanding the exponential in equation~(\ref{eq:Sqt}) to leading order in
the logarithmic contribution gives
\begin{eqnarray}\label{eq:Sqt2}
 \frac{S(q,t)}{S(q,0)} \sim \left[ 1 - \frac{q^2\Lambda^3}{\EE \ell_p
\pi^4} \left(\gamma_E + \log\frac{t}{\tau_\Lambda}\right)\right]
\exp\left[-\frac{q^2 \delta r_{\perp \Lambda}^2(\infty)}{4}\right]
\end{eqnarray}
for the unstressed test chain. As demonstrated in
figure~\ref{fig:Sqt}, the numerically evaluated dynamic structure
factor agrees with this approximation well beyond the time domain
where the logarithmic contribution in the exponent of
equation~(\ref{eq:Sqt}) is small. From the slope of the logarithmic
tails of the structure factor in a semi-logarithmic plot the
stretching parameter $\EE$ is thus immediately inferred.

\section{Microrheology}
\begin{figure}
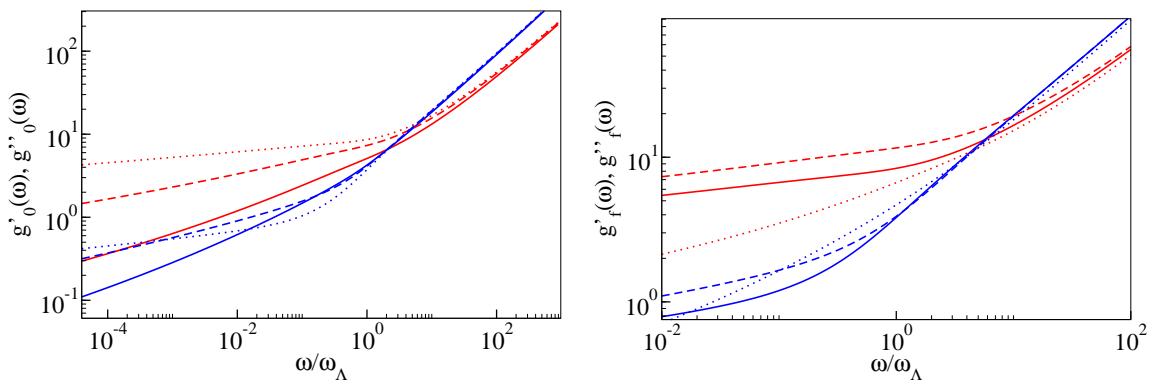

\begin{center}
\includegraphics[width=0.47\columnwidth]{gomega_trans_eps.eps}\quad
\includegraphics[width=0.47\columnwidth]{gomega_trans_eps25.eps}
\end{center}
\caption{Frequency dependence of the real and imaginary parts
 $g_f'(\omega)$, $g_f''(\omega)$ (red/blue curve) of the microrheological
 modulus, which is a common representation of the dynamic linear response to a
 transverse point force applied to a test polymer. The abscissa is
 normalised to $g_{f=0}'(\omega_\Lambda)$ ($\EE=0$).  \emph{Left:} $\EE=$ 3
 (solid), 10 (dashed), 35 (dotted) at vanishing prestress,
 $f=0$. \emph{Right:} $f/f_\Lambda=$ 0 (solid), 1 (dashed), 2.4
 (dotted) at fixed $\EE=25$ with $f_T=0.1f_\Lambda$, $f_\Lambda$ being
 the Euler force for the interaction wavelength
 $\Lambda$.}\label{fig:micro}
\end{figure}

From the transverse MSD of a point on the polymer contour, we deduce
the linear susceptibility $\alpha_f(\omega)$ (the subscript $f$ refers
to the prestressing tension) to a transverse oscillating point force
at frequency $\omega$ from the fluctuation dissipation theorem. It
relates the imaginary part $\alpha_f''(\omega)$ of the susceptibility
to the Fourier transform $\delta r_{\perp L}^2(\omega)$ of the MSD via
$\alpha_f''(\omega)=-\omega \delta r_{\perp L}^2(\omega)/2 k_BT$. The
real part of $\alpha_f$ is then uniquely determined by the
Kramers--Kronig relations, so that we find altogether
\begin{equation}
\alpha_f(\omega)=\frac{L^3}{k_BT \ell_p \pi^4 } \sum_{n=1}^\infty
\frac{1}{(n^4+n^2f/f_L)(1 +i \omega\tilde \tau_n)} \;.
\end{equation}
For better comparison with the macrorheological complex shear modulus,
it is customary to report the inverse (up to a constant scale factor)
$g^*_f(\omega)\propto 1/\alpha_f(\omega)$ of the susceptibility, which
is called the ``microrheological modulus''. Its real and imaginary
parts $g_f'(\omega)$ and $g_f''(\omega)$ are plotted in
figure~\ref{fig:micro}. The prestressing force $f$ is seen to compete
with the stretching parameter $\EE$ in raising/lowering the apparent
power-law exponent of the low-frequency modulus. Its full effect is
somewhat richer, because $f$ also affects the WLC mode amplitudes and
relaxation times according to the explicit expressions provided in
section~\ref{sec:WLC}.

\section{The shear modulus}\label{sec:modulus}
\begin{figure}
\begin{center}
\includegraphics[width=0.56\columnwidth]{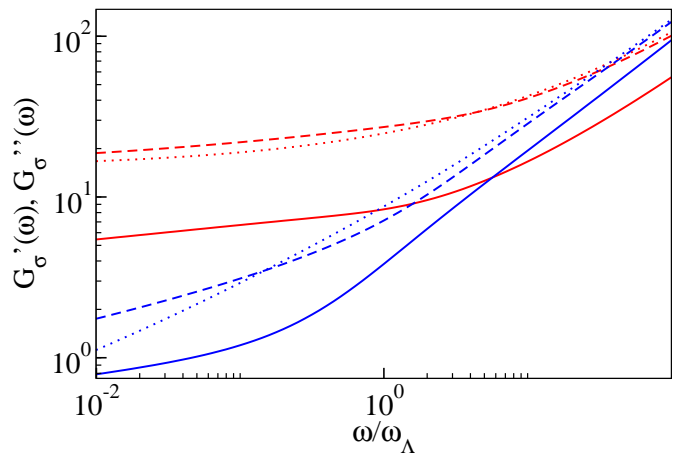}
\end{center}
\caption{Frequency dependence of the real and imaginary parts
 $G_\sigma'(\omega)$, $G_\sigma''(\omega)$ (red/blue curve) of the macroscopic
 shear modulus obtained by applying the GWLC prescription to the theoretical
 expression for the high-frequency limiting form of the shear modulus
 from Ref.~\cite{gittes-mackintosh:98_pub} for various nominal
 prestresses $\sigma=f/5\xi^2$ corresponding to $f/f_\Lambda=$ 0
 (solid), 2 (dashed), 2.45 (dotted) at $\EE=25$,
 $f_T=0.1f_\Lambda$. The modulus has been normalised to
 $G_{\sigma=0}'(\omega_\Lambda)$ ($\EE=0$).}\label{fig:shear_modulus}
\end{figure}

The microrheological modulus discussed in the preceding section should
not be confused with the shear modulus measured by a macroscopic
rheometer. While, in practice, the two quantities are sometimes hard
to distinguish, this must be attributed to non-ideal (i.e.\ not
point-like) probes such as colloidal beads used to transmit the force
to the medium and to detect its deformation, which require additional
considerations \cite{chen-etal:2003,squires-brady:2005}. Noninvasive
techniques, such as quasi-elastic light scattering are more sensitive
to the difference between the two response functions
\cite{semmrich-etal:tbp}. In the following, we present results based
on the assumption that the macroscopic shear modulus is obtained by
applying our GWLC prescription, equation~(\ref{eq:GWLC}), to the high
frequency limiting form of the shear modulus
\cite{gittes-mackintosh:98_pub}. The latter is a single polymer
quantity due to the independent relaxation of the short wavelength
modes that dominate the high frequency response. The results thus
obtained for the frequency dependence of the real and imaginary parts
$G_\sigma'(\omega)$ and $G_\sigma''(\omega)$ of the shear modulus
$G^*_\sigma(\omega)$ are displayed for various nominal prestresses
$\sigma=f/5\xi^2$ in figure~\ref{fig:shear_modulus}.  In contrast to
the microrheological modulus, the shear modulus is seen to develop a
plateau that is sensitive to the prestress as a consequence of the
underlying assumption \cite{gittes-mackintosh:98_pub} that the single
polymers are stretched affinely upon applying a macroscopic shear
stress. This assumption is presently under scrutiny
\cite{head-levine-mackintosh:2003prl,didonna-lubensky:2005,heussinger-frey:2006b},
and might in the future have to be relaxed for modes of wavelength
$\lambda_n\gtrsim \Lambda$.

\begin{figure}
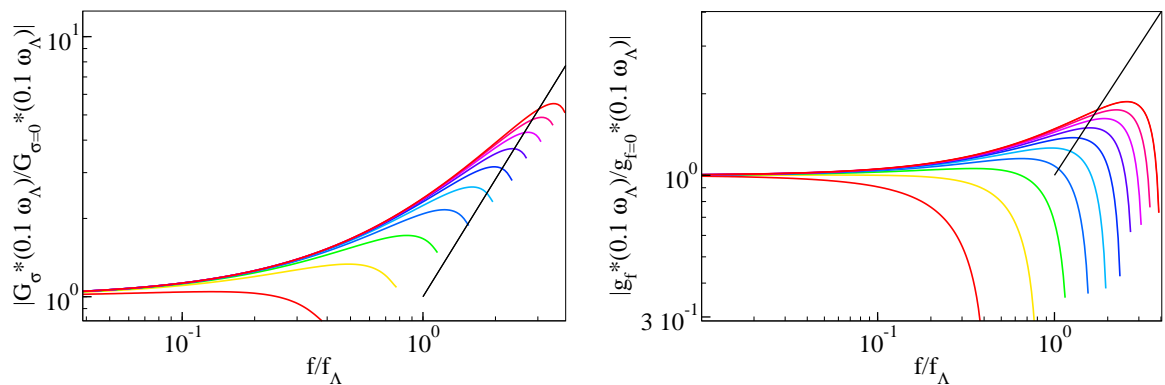

\begin{center}
\includegraphics[width=0.47\columnwidth]{hardsoft_long.eps} \quad
\includegraphics[width=0.47\columnwidth]{hardsoft_trans.eps}
\end{center}
\caption{The normalised moduli as a function of
 prestress. \emph{Left:} The shear modulus
 $|G_\sigma^*|/|G_{\sigma=0}^*|$, which is, under the assumptions
 stated in the main text, equivalent to the nonlinear differential
 shear modulus, evaluated at a fixed frequency
 $\omega=0.1\omega_\Lambda$ as a function of the normalised tension
 $f/f_\Lambda$ corresponding to a nominal prestress
 $\sigma= f/5\xi^2$ for various $\EE=4\dots 40$ (from bottom to
 top) at $f_T=0.1 f_\Lambda$. The straight line indicates the asymptotic
 stiffening power-law
 to which the stiffening curves slowly converge in the limit
 $\EE\to\infty$. \emph{Right:} The corresponding curves for transverse
 the microrheological modulus.}\label{fig:moduli}
\end{figure}

On the level of our simplifying assumptions, the shear modulus in the
presence of a prestress is equivalent to the nonlinear differential
shear modulus $K$ \cite{Gardel2004Science,gardel-etal:2004prl}. It is
therefore of some interest to evaluate the dependence of $|G^*|$ at a
fixed frequency $\omega$ as a function of the nominal prestress
$\sigma$, which then amounts to a characterisation of the nonlinear
finite-time elasticity of the system.  Via the force dependence of the
bare WLC expressions of section~\ref{sec:WLC}, the prestress causes
stress stiffening, i.e., a monotonic increase of $|G^*|$ with
$\sigma$. In contrast, the exponential speed-up of the relaxation
caused by the barrier-lowering effect of the generated tension
according to equation~(\ref{eq:kramers}) eventually overcompensates
this stiffening for large stresses. This is analysed in
figure~\ref{fig:moduli} (left) for various stretching parameters. The
limiting functional relation $|G_\sigma^*|\propto \sigma^{3/2}$ for
the stiffening, which is only slowly approached for $\EE\to\infty$ is
an immediate consequence of the underlying affine assumption
\cite{gittes-mackintosh:98_pub}.  While there is some recent
experimental support that this limiting behaviour is indeed measurable
in actin bundle networks heavily crosslinked by scruin
\cite{gardel-etal:2004prl} (plausibly corresponding to
$\EE\to\infty$), experiments for actin/$\alpha-$actinin solutions
\cite{Xu2000}, actin solutions homogeneously crosslinked by heavy
meromyosion (HMM) in the rigor state \cite{tharmann2007}, and pure actin
solutions \cite{semmrich-etal:tbp}
rather reveal a continuity of stiffening relations. They suggest that
the stiffening is much less universal than previously thought
\cite{Storm2005} and might depend on the degree of ``glassiness'' of
the system, consistent with a finite stretching parameter $\EE$
dependent on various control parameters such as crosslinker
concentration, temperature and ionic strength. To some extent,
observations of a weaker stiffening might also indicate a contribution
of (non-affine) transverse modes, as these exhibit weaker stiffening;
see the corresponding curves for the transverse microrheological
response in figure~\ref{fig:moduli} (right), which converge to the
asymptotic stiffening relation $|g_f^*|\propto f$ for very large
$\EE$. But the GWLC does neither require nor support a strict
correspondence of network affinity to the nonlinear response as
recently postulated \cite{Gardel2004Science}.

\section{Cytoskeletal networks and rough free energy landscapes}
While free energy landscapes certainly represent an appealing
intuitive framework for the discussion of many properties of complex
disordered media, it is notoriously hard to calculate their pertinent
features from first principles \cite{wales:2003}. In the remainder, we
want to give some qualitative arguments as to why we expect the rough
free energy landscapes alluded to in our motivation of the GWLC to be
characteristic of cytoskeletal networks \emph{in vitro} and \emph{in
vivo}. We distinguish two contributions to the free energy: direct
contributions from a bare polymer-polymer pair interaction potential,
and collective cageing or entanglement effects. For the typical
biopolymer solutions and networks found in cells and tissues both
cannot easily be derived microscopically or reduced to known
elementary atomic pair interactions: the former because of the
strongly collective, heterogeneous, anisotropic etc.\ nature of
protein interactions \cite{sear:2005}, the latter because semidilute
polymer solutions represent a highly correlated state of matter
\cite{schaefer:99}.

\begin{figure}
\begin{center}
\includegraphics[width=0.5\columnwidth]{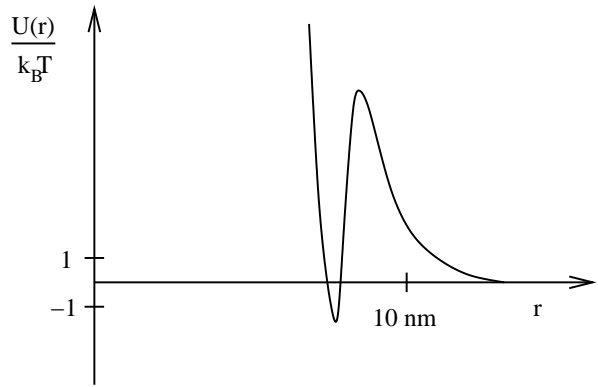}
\end{center}
\caption{Schematic sketch of a hypothetical bare pair potential $U(r)$
   suggested to provide an approximate representation of the complex
   interactions between cytoskeletal constituents.}\label{fig:pair_potential}
\end{figure}

While protein interactions remain poorly understood, many observations
hint at (unspecific, e.g. hydrophobic) adhesive contact interactions
incompletely screened by electrostatic repulsion
\cite{piazza:2004}. It is thus not implausible that direct
interactions of cytoskeletal elements can approximately be represented
by a pair potential $U(r)$ of the qualitative form sketched in
figure~\ref{fig:pair_potential} \cite{hosek-tang:2004}, which features
a narrow energy barrier.  Such ``enthalpic'' barriers slow down the
mode relaxation by an Arrhenius factor, which scales exponentially in
the barrier height. The barriers would thus yield substantial
contributions to $\EE$ without seriously affecting the thermodynamics
of the system. The less obvious free energy contributions from cageing
and entanglement are typically postulated in generic free-volume
theories \cite{stinchcombe-depken:2002}, as their calculation from
first principles remains one of the major challenges in the theory of
structural glasses \cite{gujrati-rane-corsi:2003}. For semiflexible
polymer networks they can be estimated within the tube model
\cite{kroy:2006}, which suggests a contribution $\EE\simeq
1$. Accordingly, a plausible estimate for the effective well width
$\Delta$ should be given by the tube diameter, which scales like
$\xi^{6/5}\ell_p^{-1/5}$ \cite{semenov:86} in a semidilute
semiflexible polymer solution and is estimated to assume values in the
range of some 10 to 100 nm for typical \emph{in vitro} polymerised
actin solutions. In any case, simple exponential scaling of the
relaxation times in $\EE$ and in the wavelength $\lambda$, as
postulated in equation~(\ref{eq:GWLC}) seems plausible for an
entangled solution of weakly bending sticky polymers.

It is important to realise that the parameters $\EE$ and $\Delta$ are
to be understood as effective parameters. They cannot generally be
expected to correspond directly to some salient features of a bare
interaction potential as sketched in figure~\ref{fig:pair_potential}
nor to the experimentally more accessible coarse-grained interactions
between adjacent polymer segments of length $\Lambda$.  (The latter
can be roughly thought of as a smeared-out version of the former, due
to the effect of thermal undulations of the polymers and possible
compliant crosslinkers \cite{Wagner2006}.)

The above suggestion to interpret $\EE$ essentially as a kind of
kinetic ``stickiness'' parameter for cytoskeletal polymers might have
interesting implications as to the physiological role played by the
broad class of actin binding proteins such as crosslinkers and
molecular motors. While little is known about the microscopic
interactions between cytoskeletal polymers, even less is known about
how they might be affected by these proteins. Yet it seems plausible
that the predominant effect of most crosslinkers can be subsumed into the
parameter $\EE$. Also, while current
approaches to the active rheology of cells
\cite{hatwalne-etal:2004,kruse-etal:2004,shen-wolynes:2005} put much
emphasis on the dynamic role of molecular motors in causing transport
and motion, some of them might well play a less flamboyant role most
of the time: namely adjusting the effective stickiness $\EE$ and
tension $\sigma$ to tune the viscoelastic properties of the
cytoskeleton while saving it from glassy arrest
\cite{humphrey-etal:2002}.

\section{Conclusion}
In summary, we have presented a simple modification of the standard
model of a semiflexible polymer (the WLC), which we call the glassy
wormlike chain (GWLC), and which is obtained by an exponential
stretching of the relaxation spectrum of the WLC. The striking
resemblance of the predicted mechanical observables with rheological
data for cells and reconstituted cytoskeletal model systems naturally
suggests that these systems must exhibit strong rheological
redundancy. In fact, the perplexing universality and robustness of the
mechanical performance of biological cells and tissues against
structural modifications by drugs and mutations has been an enigma in
cell biology for quite some time \cite{sackmann:97,Fabry2001}. Our
glassy wormlike chain model offers a very economical explanation in
terms of the difference between the stretching parameter $\EE$ and the
prestress $\sigma$ (in natural units). If changes in the structure and
interactions of the cytoskeleton affect its rheological properties
chiefly via $\EE$ and $\sigma$, the potential for mutual compensation
appears indeed enormous.  This leads us to the suggestion that a
microscopic calculation of the various anticipated contributions to
$\EE$ will be one of the major challenges for future theoretical work
aiming to explain the rheology of cells and \emph{in vitro}
reconstituted models of the cytoskeleton.

\section*{References}

\bibliographystyle{prsty}

\end{document}